# Probabilistic Greedy Algorithm Solver Using Magnetic Tunneling Junctions for Traveling Salesman Problem

**Thomas Kämpfe**

thomas.kaempfe@ipms.fraunhofer.de

Fraunhofer IPMS   https://orcid.org/0000-0002-4672-8676

**Ran Zhang**

Beijing National Laboratory for Condensed Matter Physics, Institute of Physics, University of Chinese Academy of Sciences, Chinese Academy of Sciences   https://orcid.org/0000-0001-5711-6475

**Xiaohan Li**

Beijing National Laboratory for Condensed Matter Physics, Institute of Physics, University of Chinese Academy of Sciences, Chinese Academy of Sciences, Beijing 100190, China

**Cai-Hua Wan**

Chinese Academy of Sciences   https://orcid.org/0000-0002-5456-1052

**Raik Hoffmann**

Fraunhofer IPMS

**Meike Hindenberg**

Fraunhofer IPMS

**Yingqian Xu**

Beijing National Laboratory for Condensed Matter Physics, Institute of Physics, University of Chinese Academy of Sciences, Chinese Academy of Sciences

**Shiqiang Liu**

Beijing National Laboratory for Condensed Matter Physics, Institute of Physics, University of Chinese Academy of Sciences, Chinese Academy of Sciences, Beijing 100190, China

**Dehao Kong**

Beijing National Laboratory for Condensed Matter Physics, Institute of Physics, University of Chinese Academy of Sciences, Chinese Academy of Sciences

**Shilong Xiong**

Beijing National Laboratory for Condensed Matter Physics, Institute of Physics, University of Chinese Academy of Sciences, Chinese Academy of Sciences

**Shikun He**

Zhejiang Hikstor Technology Co. LTD.

**Alptekin Vardar**

Fraunhofer IPMS

**Qiang Dai**
Zhejiang Hikstor Technology Co. Ltd

**Junlu Gong**
Zhejiang Hikstor Technology Co. Ltd

**Yihui Sun**
Zhejiang Hikstor Technology Co. Ltd

**Zejie Zheng**
Zhejiang Hikstor Technology Co. Ltd

**Guoqiang Yu**
Beijing National Laboratory for Condensed Matter Physics, Institute of Physics, Chinese Academy of Sciences, Beijing 100190, China    https://orcid.org/0000-0002-7439-6920

**Xiufeng Han**
Beijing National Laboratory for Condensed Matter Physics, Institute of Physics, University of Chinese Academy of Sciences, Chinese Academy of Sciences, Beijing 100190, China    https://orcid.org/0000-0001-8053-793X

**Article**

**Additional Declarations:** There is **NO** Competing Interest.

# Probabilistic Greedy Algorithm Solver Using Magnetic Tunneling Junctions for Traveling Salesman Problem

Ran Zhang[1], Xiaohan Li[1], Caihua Wan[1, 2, 6, *], Raik Hoffmann[3], Meike Hindenberg[3], Yingqian Xu[1], Shiqiang Liu[1], Dehao Kong[1], Shilong Xiong[1], Shikun He[4], Alptekin Vardar[3], Qiang Dai[4], Junlu Gong[4], Yihui Sun[4], Zejie Zheng[4], Thomas Kämpfe[3, 5, *], Guoqiang Yu[1, 2, 6] and Xiufeng Han[1, 2, 6, *]

[1]Beijing National Laboratory for Condensed Matter Physics, Institute of Physics, University of Chinese Academy of Sciences, Chinese Academy of Sciences, Beijing 100190, China

[2]Center of Materials Science and Optoelectronics Engineering, University of Chinese Academy of Sciences, Beijing 100049, China

[3]Fraunhofer IPMS, Center Nanoelectronic Technologies, 01109 Dresden, Germany

[4]Zhejiang Hikstor Technology Co. Ltd, Hangzhou 311305, China

[5]TU Braunschweig, Institute for CMOS Design, 38106 Braunschweig, Germany

[6]Songshan Lake Materials Laboratory, Dongguan, Guangdong 523808, China

E-Mail: thomas.kaempfe@ipms.fraunhofer.de ; wancaihua@iphy.ac.cn ; xfhan@iphy.ac.cn

**Combinatorial optimization problems are foundational challenges in fields such as artificial intelligence, logistics, and network design. Traditional algorithms, including greedy methods and dynamic programming, often struggle to balance computational efficiency and solution quality, particularly as problem complexity scales. To overcome these limitations, we propose a novel and efficient probabilistic optimization framework that integrates true random number generators (TRNGs) based on spin-transfer torque magnetic tunneling junctions (STT-MTJs). The inherent stochastic switching behavior of STT-MTJs enables dynamic configurability of random number distributions, which we leverage to introduce controlled randomness into a probabilistic greedy algorithm. By tuning a temperature parameter, our algorithm seamlessly transitions between deterministic and stochastic strategies, effectively balancing exploration and exploitation. Furthermore, we apply this framework to the traveling salesman problem (TSP), showcasing its ability to consistently produce high-quality solutions across diverse problem scales. Our algorithm demonstrates superior performance in both solution quality and convergence speed compared to classical approaches, such as simulated annealing and genetic algorithms. Specifically, in larger TSP instances involving up to 70**

**cities, it retains its performance advantage, achieving near-optimal solutions with fewer iterations and reduced computational costs. This work highlights the potential of integrating MTJ-based TRNGs into optimization algorithms, paving the way for future applications in probabilistic computing and hardware-accelerated optimization.**

## 1. Introduction

Combinatorial optimization is a cornerstone of modern computational science, playing a pivotal role in domains ranging from artificial intelligence and machine learning[1,2] to logistics[3-5] and operations research[6]. The objective is to identify an optimal configuration from a finite but exponentially large set of possibilities, where even modest increases in problem size can render classical methods impractical due to the exponential growth of computational complexity[7]. While deterministic algorithms such as dynamic programming[8] and branch-and-bound[9] have proven effective for small-scale problems, they often fail to scale efficiently to larger scenarios or escape local optima when confronted with the complex landscapes of combinatorial spaces[10].

In recent years, there has been a paradigm shift towards incorporating randomness into optimization algorithms[11-13], leading to a new class of techniques termed stochastic or probabilistic optimization[14-20]. Methods such as simulated annealing, genetic algorithms, and Monte Carlo simulations have demonstrated the potential of randomness to diversify search strategies, enabling algorithms to explore solution spaces more comprehensively and escape local minima. However, the efficacy of these methods is highly dependent on the quality and configurability of the random number generators (RNGs) employed[13,21,22]. Traditional RNGs, whether pseudo-random or hardware-based, often lack the flexibility required to dynamically adjust their distribution characteristics, limiting their adaptability to different optimization scenarios.

A promising development in this field is the utilization of magnetic tunneling junctions (MTJs) as a source of true random numbers[23,24]. MTJs, typically used in non-volatile memory technologies[25-28], exhibit probabilistic switching behavior that can be finely tuned by external control parameters such as voltage or magnetic field strength. This inherent stochasticity makes MTJ-based true random number generators (TRNGs) uniquely suited for probabilistic computing[29,30], where the randomness can be directly mapped onto computational processes. The ability to configure the probability distribution of an MTJ-based TRNG enables a novel approach to algorithm design, where the degree of randomness can be adjusted in real time to influence decision-making processes[31,32].

In this study, we propose an advanced optimization framework that leverages MTJ-based TRNGs to solve complex combinatorial problems. Specifically, we introduce a **probabilistic greedy algorithm** for the traveling salesman problem (TSP)[22,33-36] – a canonical example in combinatorial optimization – to showcase the potential of this approach. The TSP challenges a solver to find the shortest possible route that visits a given set of cities and returns to the starting point, and it is well known for its non-deterministic polynomial-hardness (NP-Hardness). By incorporating MTJ-based TRNGs into the decision-making process, we can modulate the selection strategy for the next city, transitioning smoothly between deterministic greedy choices and purely random selection. This dynamic adaptability enables the algorithm to effectively balance exploration and exploitation, thereby improving its ability to find high-quality solutions efficiently.

## 2. Methods:

The stack structure of the employed STT-MTJ devices[37-39], as depicted in Figure 1a, is from top to bottom capping/CoFeB/Mo/CoFeB/MgO/CoFeB/Mo/[Co/Pt]$_n$-based synthetic anti-ferromagnetic structure/Seed/SiO$_2$. The multilayer films were deposited by magnetron sputtering on a thermally oxidized silicon substrate under a vacuum environment of $10^{-6}$ Pa. Following deposition, the films were annealed at high temperature in an external magnetic field perpendicular to the film plane. The devices were then patterned into cylindrical STT-MTJs using standard lithography and etching processes. Magneto-transport measurements of the fabricated devices were conducted using an Hprobe H3DM tester. The samples were subsequently connected to a Keysight B1500A semiconductor analyzer and a NI PXIe system through a probe card and adapter board, enabling comprehensive experimental control and data acquisition through a Python-based interface. This setup facilitated precise electrical measurements and switching probability characterization of the STT-MTJs, providing a reliable platform for evaluating their performance as TRNGs.

## 3. Results and Discussion:

Figure 1 presents a detailed characterization of the MTJ-based TRNG employed in this study. The resulting *R-H* hysteresis loops, shown in Figure 1b, reveals a clear and sharp switching between high and low resistance states, confirming the stability and reproducibility of the MTJ's magnetic switching behavior. The MTJ has a high tunnel magnetoresistance (TMR) ratio ~175%, which is essential for ensuring reliable and distinct resistance states. The MTJ's resistance switching behavior under current pulses is illustrated in Figure 1c.

By applying a series of current pulses, we observed stochastic switching of the free layer's magnetization, resulting in resistance changes. This stochastic behavior serves as the basis for the MTJ-based TRNG. To further analyze the switching probability, Figure 1d plots the probability of switching as a function of the applied write voltage. The experimental data (blue circles) show a gradual increase in switching probability ($P_{sw}$) with increasing voltage ($V$), which is accurately captured by the fitted sigmoidal curve (black solid line). The fitting parameters $b$ and $c$ indicate the sharpness and offset of the curve.

$$P_{sw} = \frac{1}{1+e^{-b(V+c)}}$$

This relationship $P_{sw}(V)$ is crucial, as it enables precise control over the probability distribution of generated random numbers. Figure 1e demonstrates the results of continuous resistance measurements at three fixed voltages: 0.275 V, 0.282 V, and 0.288 V, corresponding to $P_{sw}$ of 25%, 48%, and 81%, respectively. These measurements confirm that the device can achieve consistent and repeatable switching behavior, with well-defined probability at each voltage level. The ability to finely tune the switching probability by adjusting the voltage is a key advantage of MTJ-based TRNGs, allowing for the generation of random numbers with specific statistical properties tailored to various probabilistic algorithms.

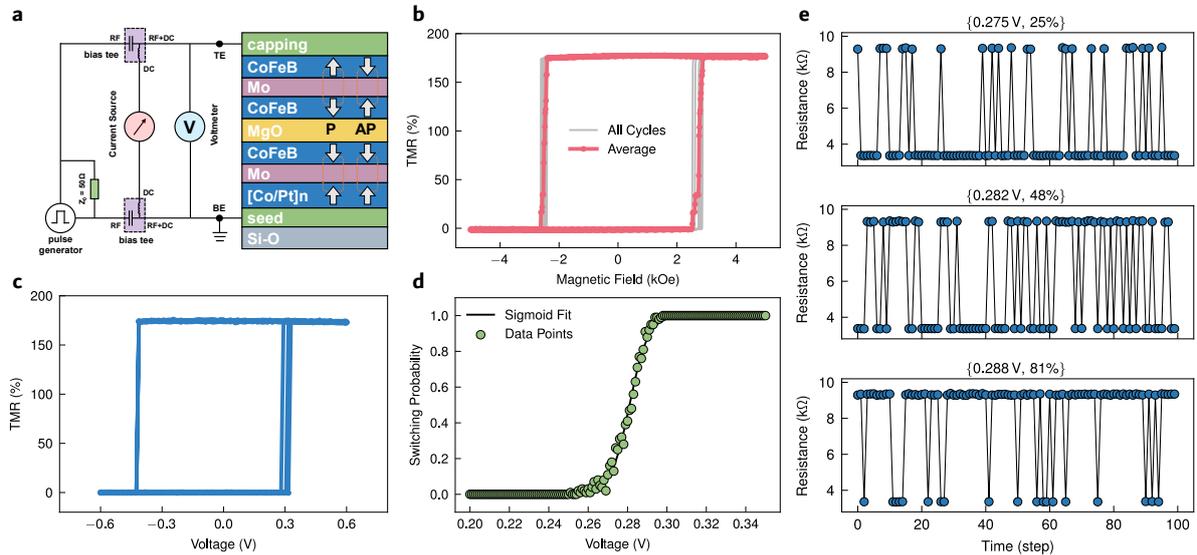

**Figure 1. Characterization of the performance of a TRNG based on an MTJ. a**, Schematic of the device structure and measurement setup. **b**, *R-H* hysteresis loops obtained by sweeping an out-of-plane magnetic field. **c**, Resistance switching behavior of the MTJ induced by current pulses, which trigger free layer magnetization switching. **d**, MTJ switching probability as a function of applied write voltage. The black solid line represents the fitted sigmoidal curve. **e**, Resistance measurements from continuous testing

at fixed voltages of 0.275 V, 0.282 V, and 0.288 V, corresponding to switching probabilities of 25%, 48%, and 81%, respectively.

Figure 2 showcases the versatility of the MTJ-based TRNG in generating random numbers with configurable probability distributions. The schematic diagram in Figure 2a illustrates the experimental setup, where multiple MTJs are connected to the NI PXIe system through a probe card and adapter board (Supplement I). This configuration enables the simultaneous measurement of multiple MTJs, allowing for efficient data collection and parallel testing of different devices.

Figure 2b displays the random numbers generated by the MTJ-based TRNGs. The generated values align closely with the expected Gaussian distribution, as evidenced by the smooth bell-shaped curve. This Gaussian-distributed randomness is achieved by carefully adjusting the write voltage of the MTJs, demonstrating the flexibility of the TRNG in producing specific distributions. The transformation from binary Bernoulli TRNGs into a probability-distribution-function-configurable TRNG modeled as a Bayesian network can be found in Supplement II in details.

To quantitatively evaluate the accuracy of the generated distributions, Figure 2c presents the error analysis, where the left axis represents the Kullback-Leibler (KL) divergence and the right axis represents the mean squared error (MSE). The KL divergence measures the difference between the experimentally generated distribution and the theoretical Gaussian distribution, while the MSE quantifies the average deviation of the generated values from the expected mean and variance. Both metrics indicate minimal errors, confirming the high fidelity of the MTJ-based TRNG in replicating desired distributions.

The neighbor correlation of the generated random numbers is analyzed in Figure 2d, where the color intensity represents the sample point density. The nearly uniform distribution of points and the presence of concentric circles indicate negligibly weak neighboring correlation, signifying that the generated random numbers are statistically independent. Our STT-MTJs are not low-barrier ones and each random number is generated by a reset-sampling circle, therefore correlevance between neighboring random numbers no longer an issue here. This property is essential for ensuring that the TRNG can produce high-quality random numbers suitable for applications requiring true randomness, such as probabilistic algorithms and cryptographic operations.

Figures 2e to 2g demonstrate the capability of the TRNG to generate random numbers following various probability distributions. Figure 2e presents a uniform distribution, where each value has an equal probability of being sampled. Figure 2f shows an exponential decay distribution, characterized by a high probability for smaller values and a rapidly decreasing probability for larger values. Finally, Figure 2g illustrates a user-defined arbitrary distribution, highlighting the flexibility of the TRNG in generating custom probability profiles. This configurability is critical for integrating the TRNG into a wide range of applications, from stochastic optimization to artificial intelligence, where diverse probability distributions are needed to guide decision-making processes.

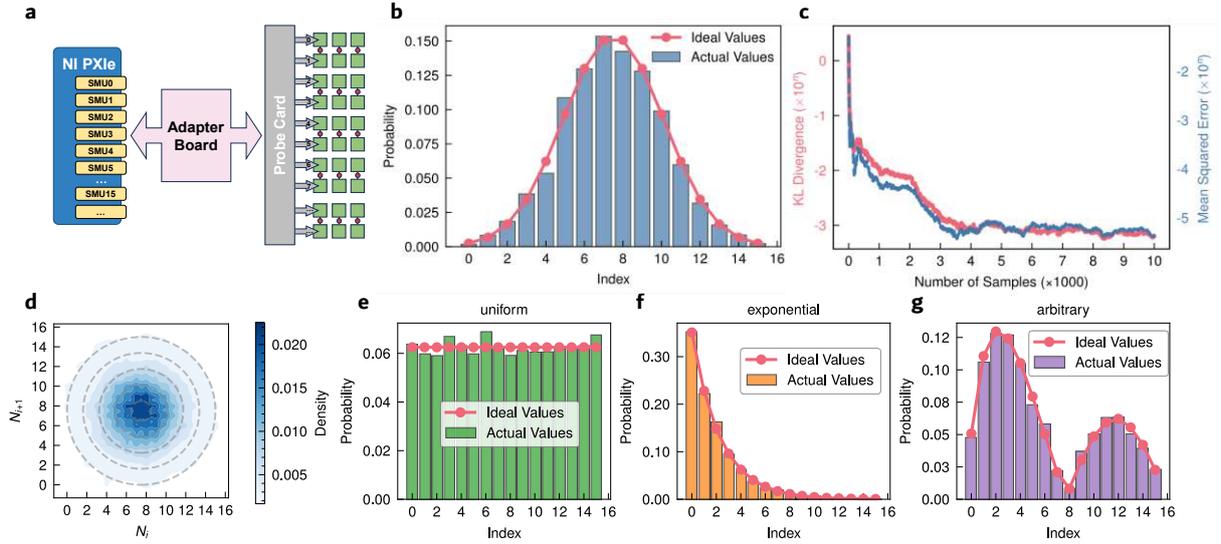

**Figure 2. Probability-distribution-configurable TRNGs based on multiple MTJs. a**, Schematic diagram illustrating the connection of multiple MTJs to the NI PXIe system through a probe card and adapter board. **b**, Generated random numbers exhibiting a Gaussian distribution. **c**, Error analysis of the Gaussian-distributed random numbers, with the left axis representing KL divergence and the right axis representing the mean squared error. **d**, Neighbor correlation of the Gaussian-distributed random numbers. The color intensity indicates the sample point density, and the concentric circles indicate weak neighbor correlation. **e-g**, Random number generation results for several typical probability distributions: (**e**) uniform distribution, (**f**) exponential decay distribution, and (**g**) user-defined arbitrary distribution.

Figure 3 presents the probabilistic greedy algorithm's mechanism for selecting the next city in the traveling salesman problem (TSP) under varying temperature conditions. The algorithm utilizes the MTJ-based TRNG to generate random numbers that influence the city selection process, allowing for a probabilistic adjustment of the greedy strategy. The selection probability $P_{i+1}(\bar{N})$ of the next city $\bar{N}$ is a function of the distance $d_{ij}$ between the current city $N$ and $\bar{N}$ as well as a temperature parameter $k_B T$ as shown in Eq. (1).

$$P_{i+1}(\bar{N}_i) = (1-b_i)\exp(-d_{N\bar{N}_i}/k_B T)/Z$$
$$Z = \sum_{i=1}^{8}(1-b_i)\exp(-d_{N\bar{N}_i}/k_B T)$$
(1)

Here $b_i$ indicates the accessibility of the $i^{th}$ city and $b_i=1$ once the $i^{th}$ city has been visited or else $b_i=0$ if it is to be visited. Thus, the final probability of choosing a specific route $P=\prod P_{i+1}(\overline{N}) \propto \exp[-(\sum d_{ij})/k_B T]$ and, straightforwardly, the shortest route $S = (\sum d_i)_{min}$ has the highest probability to be experimentally sampled. This feature assures the convergence of this probabilistic greedy algorithm. More details can be found in the Supplement III.

It is worth noting that the decision of choosing the next city relies on a probabilistic sampling operation according to the series of probabilities $P_{i+1}(\overline{N})$ with $\overline{N}$ being the city indices to be visited. This probability-distribution-function (PDF) defined by $P_{i+1}(\overline{N})$ changes dynamically step by step, which calls for a random number generator that can output random numbers according to the time-variant PDFs. Fortunately, our TRNGs with configurable PDFs match this requirement well.

When $k_B T$ approaches zero, the algorithm operates as a deterministic greedy algorithm, always selecting the closest city to the current one. In this regime, the probability of choosing the closest city is nearly 100%, leading to rapid but potentially suboptimal solutions due to the algorithm's inability to escape local minima.

Conversely, when $k_B T$ is extremely high, the selection probability for each remaining city becomes nearly uniform, leading to a selection process similar to a random walk. This behavior encourages exploration of the solution space but at the cost of reduced efficiency in converging to high-quality solutions. The optimal performance is observed at intermediate $k_B T$-values, where the algorithm effectively balances exploration (randomness) and exploitation (favoring shorter distances), allowing it to escape local optima and discover near-optimal solutions with high probability.

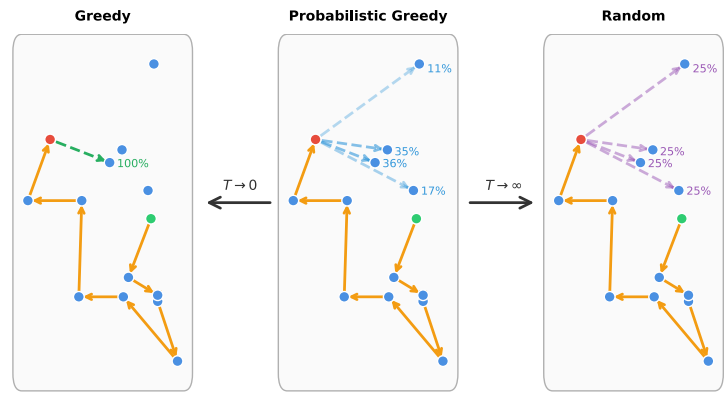

**Figure 3. Probability of selecting the next city under different temperature ($T$) conditions.** When $T$ approaches 0, the algorithm always selects the closest city, equivalent to a greedy algorithm. When $T$ approaches infinity, the selection probability for each remaining city is equal,

which is analogous to random selection. When $T$ is within a suitable intermediate range, a probabilistic greedy algorithm can be achieved.

Figure 4 provides experimental results demonstrating the application of the MTJ-based TRNGs in solving the TSP using the probabilistic greedy algorithm. Figure 4a depicts the map of the Burma14 TSP problem, where the solid line indicates the known optimal solution, and the dashed line represents the best solution obtained using a classic greedy algorithm. The probabilistic greedy algorithm, driven by the MTJ-based TRNG, consistently identifies paths that are closer to the optimal solution, as shown by the reduced total distance metrics. Figure 4b illustrates the variation in total distance across a range of $k_BT$ values from 1 to 400. The orange dashed line marks the known optimal solution, while the red, green, and orange solid lines connect the maximum, minimum, and average total distances, respectively, obtained at each $k_BT$ value. The results indicate that the algorithm achieves optimal or near-optimal solutions when $k_BT$ is within the range of 40 to 60, highlighting the significance of selecting an appropriate temperature parameter to balance the probabilistic selection strategy.

Figure 4c further investigates the distribution of solution distances at six selected $k_BT$ values, showing improved performance and reaching the optimal solution when $k_BT$ is between 40 and 60. This analysis underscores the robustness of the probabilistic greedy algorithm in finding high-quality solutions when driven by suitably tuned randomness. Figure 4d examines the relationship between the best path distance and the number of repetitions for four selected $k_BT$ values. When $k_BT=60$, the optimal solution is achieved within 1000 repetitions, demonstrating the efficiency of the algorithm in converging to high-quality solutions. Figure 4e presents a scatter plot of the best solutions obtained across the $k_BT$ range and density distribution plots of solutions within 0, 50 and 100 kilometers of the known optimal solution, further validating the algorithm's effectiveness. The appearance of a clear peak shape in the distribution plots indicates the existence of an optimal $k_BT$, highlighting the algorithm's sensitivity to temperature parameters in achieving high-quality solutions.

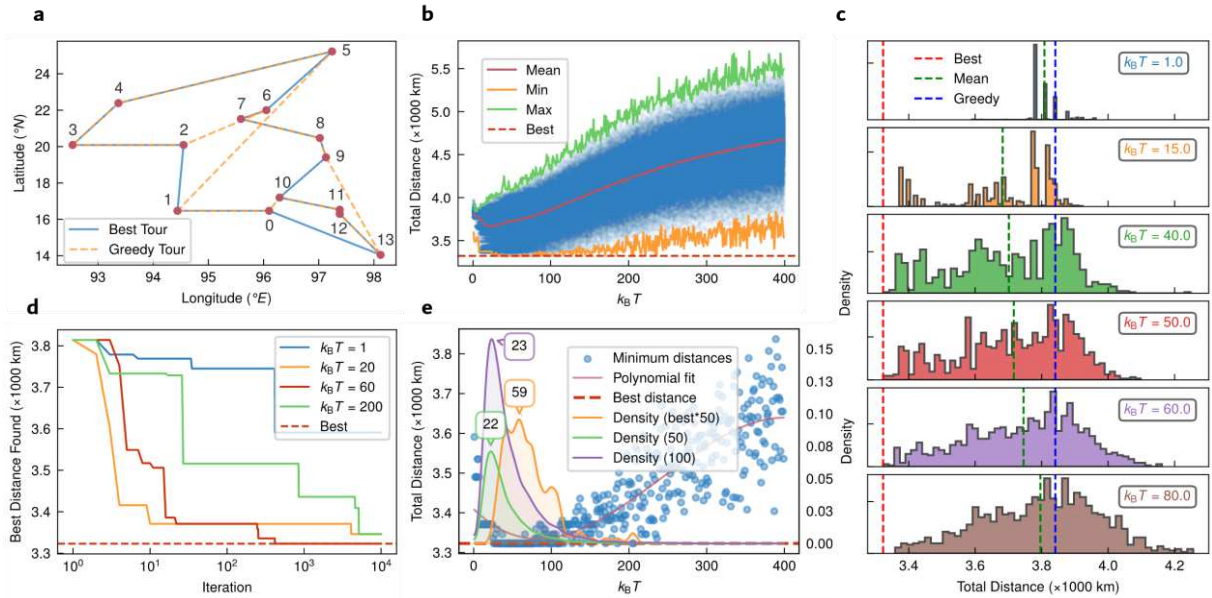

**Figure 4. Hardware test results for solving the Burma14 problem using the Probabilistic Greedy Algorithm. a**, Map of the Burma14 problem, where the solid line represents the known optimal solution and the dashed line indicates the best solution obtained using the classic greedy algorithm. **b**, Total distance statistics of TSP solutions across the range of $k_BT$=1 to 400. The orange dashed line marks the known optimal solution, while the red, green, and orange solid lines connect the maximum, minimum, and average total distances, respectively, obtained at each $k_BT$ value. **c**, Distribution of solution distances at six selected $k_BT$ values, showing improved performance and reaching the optimal solution when $k_BT$ is in the range of 40 to 60. **d**, Relationship between the best path distance and the number of repetitions for four selected $k_BT$ values. When $k_BT$=60, the optimal solution can be achieved within 1000 repetitions. **e**, Scatter plot of the best solutions across $k_BT$=1 to 400 (left) and density distribution plots of solutions within 0, 50, and 100 kilometers of the known best solution (right).

The solver still works well when the city number is increased significantly. Figure 5 illustrates the simulated results obtained with the st70 problem, offering a comprehensive comparison of our algorithm against other established methods. In the map shown for the st70 problem, the optimal solution is indicated by a solid line, serving as a benchmark for evaluating the performance of different algorithms. The st70 problem, with its 70 cities, presents a considerable computational challenge, making it an ideal test case for demonstrating the efficacy of both heuristic and exact algorithms. The comparison of time and space complexity among the algorithms—Brute Force, Dynamic Programming, Genetic Algorithm, Simulated Annealing (SA), Greedy Algorithm, and Probabilistic Greedy Algorithm—clearly highlights the benefits of heuristic methods. While exhaustive approaches like Brute Force and Dynamic

Programming struggle with scalability as $n$ increases, heuristic algorithms, particularly the Probabilistic Greedy and Genetic algorithms, strike a balance between computational efficiency and solution quality. This contrast is evident from the results where $n = 70$ is used, showcasing the advantage of these more advanced approaches when tackling larger problems. As the number of iterations grows, the quality of the solutions improves, particularly when varying the thermal fluctuation parameter $k_BT$. Across different values ($k_BT = 0.1, 1.3, 2.0$, and $3.0$), the results indicate that the algorithm's performance is highly sensitive to this parameter, with intermediate values (e.g., 1.3) leading to a more optimal convergence rate. The gradual improvement in path quality with increasing iterations underscores the algorithm's ability to refine its solution over time. When comparing different heuristic approaches—Genetic Algorithm, Simulated Annealing (SA), Greedy Algorithm, and Probabilistic Greedy Algorithm—the results reveal that incorporating stochastic elements, as seen in the Probabilistic Greedy Algorithm, significantly enhances performance. It is worth noting that while the SA can only evaluate the situation of exchanging two cities at a sample, the Probabilistic Greedy Algorithm can take all the remaining cities into account for a single sampling owing to the arbitrary PDF configurability of our MTJ-TRNGs. Thus, the latter can deal with a higher entanglement degree, which accounts for its faster convergence speed. By avoiding local optima, the probabilistic variant consistently outperforms the classic Greedy Algorithm, especially in later iterations, demonstrating its potential for yielding superior solutions.

Finally, the schematic in Figure 5e illustrates the core design of a TSP solver based on an MTJ array. This hardware-based solver taps into the inherent randomness of the MTJ array, which can be efficiently reused to generate random numbers of any required length. This modularity and scalability make the MTJ array particularly well-suited for probabilistic algorithms like Simulated Annealing and Probabilistic Greedy Algorithm, enhancing the solver's adaptability for larger and more complex TSP instances. The ability to expand the random number generation capability of the MTJ array according to an arbitrarily customized PDF without sacrificing performance is a crucial innovation, positioning this design as a versatile and efficient solution for hardware-accelerated optimization tasks.

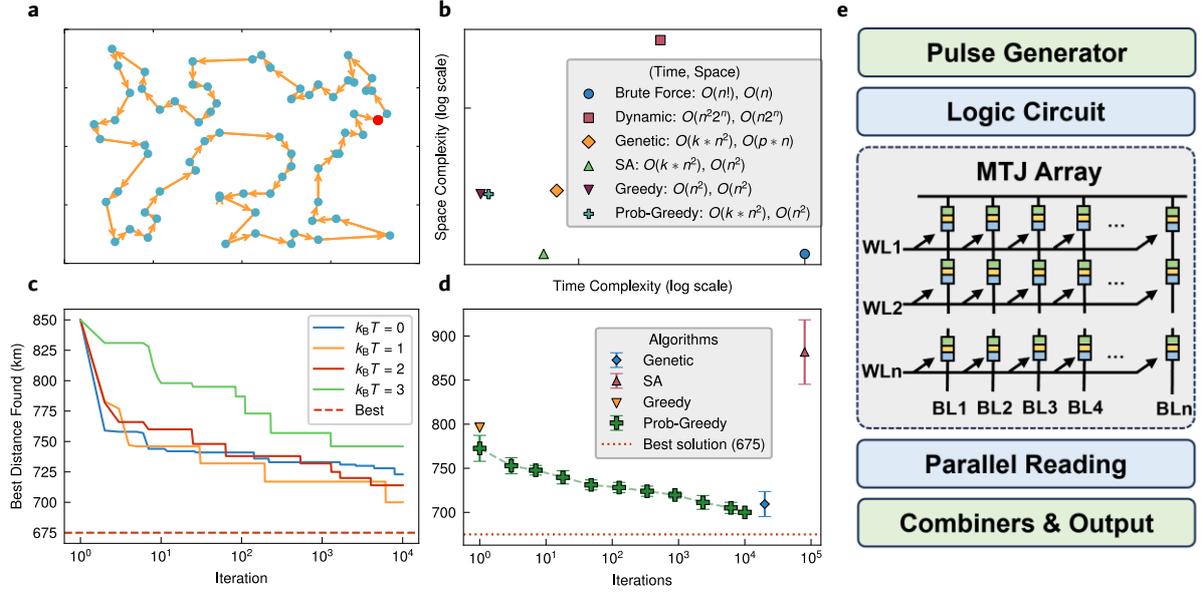

**Figure 5. Simulated results with more cities and comparison with other algorithms. a**, Map of the st70 problem, where the solid line represents the known optimal solution. **b**, A comparison of time complexity and space complexity among Brute Force, Dynamic Programming algorithm, Genetic algorithm, Simulated Annealing (SA) algorithm, Greedy algorithm, and Probabilistic Greedy algorithm, using values for *n*=70. **c**, Illustration of how the best path found in the st70 problem decreases as the number of iterations increases for $k_BT$ = 0.1, 1.3, 2.0, and 3.0. **d**, Comparison of solutions found by heuristic algorithms (Genetic algorithm, SA algorithm, Greedy algorithm, and Probabilistic Greedy algorithm), with the horizontal axis representing the number of iterations. **e**, A schematic diagram of the core design of a TSP solver based on an MTJ array, highlighting the key components and connections within the solver architecture.

## 4. Conclusion:

This paper presents a novel probabilistic greedy algorithm that utilizes the stochastic properties of MTJ-based TRNGs to solve complex combinatorial optimization problems. By integrating MTJ-based TRNGs with the PDF reconfigurability into the optimization framework, we can dynamically adjust the degree of randomness in the decision-making process, allowing the algorithm to strike an optimal balance between exploration and exploitation. This capability is achieved through the control of a temperature parameter, which modulates the randomness level and enables the algorithm to adapt its strategy based on the problem state.

The effectiveness of the proposed approach is demonstrated through extensive experimentation on the traveling salesman problem. Our results show that the probabilistic greedy algorithm consistently achieves superior performance compared to classical methods such as simulated annealing and genetic algorithms, both in terms of solution quality and convergence speed. When applied to larger problem instances in simulation, the algorithm exhibits excellent scalability and robustness, maintaining a competitive edge even as the number of cities increases to 70. The key advantage of this approach lies in its ability to dynamically modulate randomness through the MTJ-based TRNG, which enhances the algorithm's capacity to escape local optima and discover near-optimal solutions efficiently.

The integration of MTJ-based TRNGs offers a promising direction for developing hardware-accelerated optimization frameworks, with potential applications extending beyond TSP to other NP-hard problems. Future work will explore the integration of these TRNGs into parallel and distributed computing architectures, as well as their combination with advanced machine learning models to further expand the capabilities of probabilistic optimization methods. This research establishes a solid foundation for leveraging hardware-level stochasticity in computational algorithms, offering new possibilities for tackling complex optimization challenges with greater efficiency and effectiveness[4].

**Supporting Information**

Supporting Information is available from the Nature Communication or from the author.

**Acknowledgements**

This work was supported by the National Key Research and Development Program of China (MOST) (Grant No. 2022YFA1402800), the National Natural Science Foundation of China (NSFC) (Grant Nos. 12134017, 51831012, 51620105004, and 12374131), the Strategic Priority Research Program (B) of Chinese Academy of Sciences (CAS) (Grant Nos. XDB33000000). C. H. Wan appreciates financial support from the Youth Innovation Promotion Association, CAS (Grant No. 2020008).

**Conflicts of Interests**:

The authors declare no competing interests.

**Data Availability**:

All data needed to evaluate the conclusions in the paper are present in the paper. Additional data available from authors upon request.

# Figures

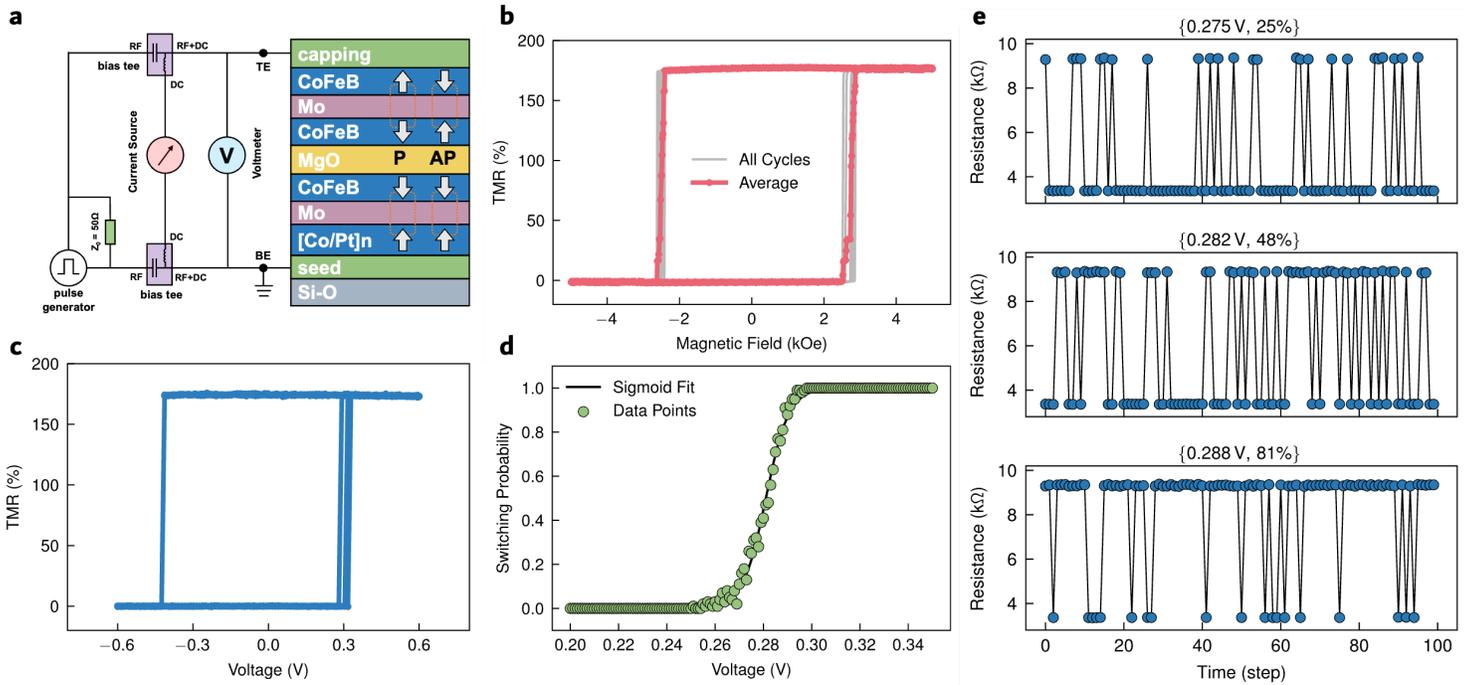

## Figure 4

showcases the versatility of the MTJ-based TRNG in generating random numbers with configurable probability distributions. The schematic diagram in Fig. a illustrates the experimental setup, where multiple MTJs are connected to the NI PXIe system through a probe card and adapter board (Supplement I). This configuration enables the simultaneous measurement of multiple MTJs, allowing for efficient data collection and parallel testing of different devices.

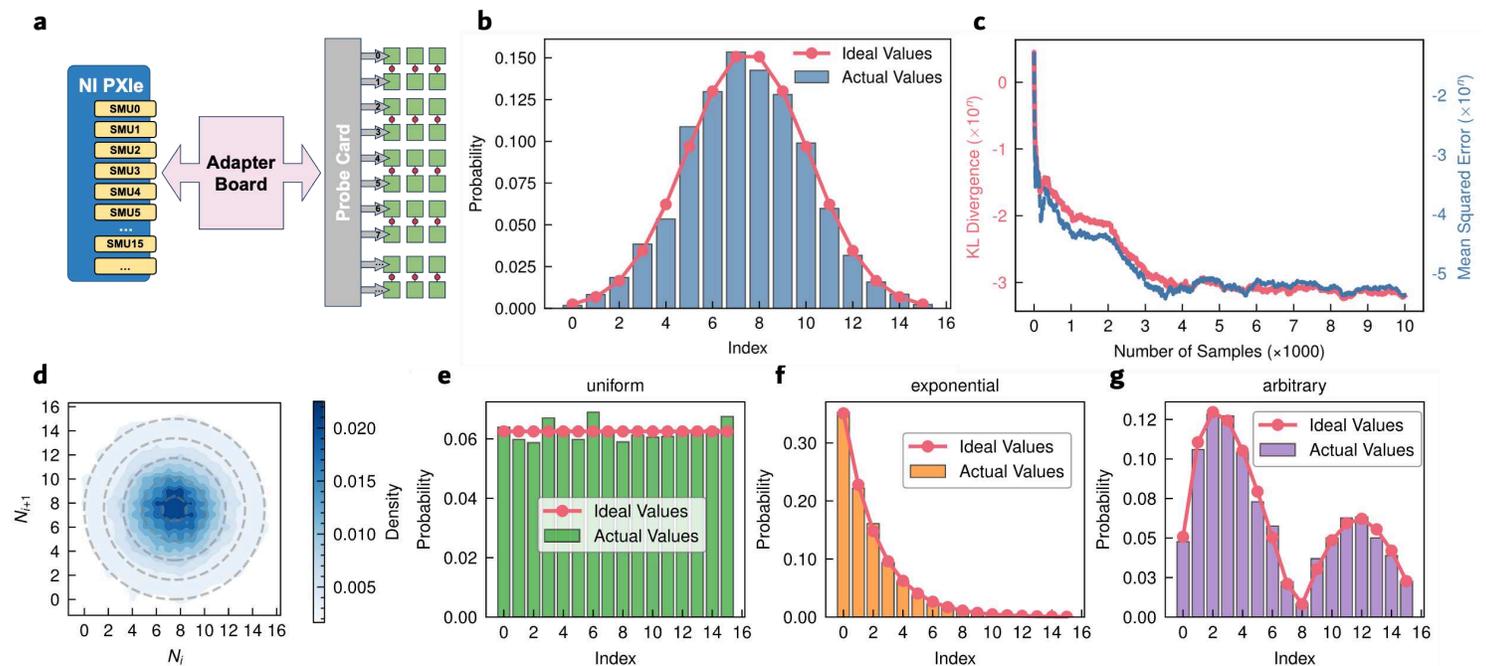

## Figure 5

to 2g demonstrate the capability of the TRNG to generate random numbers following various probability distributions. Figure e presents a uniform distribution, where each value has an equal probability of being sampled. Figure f shows an exponential decay distribution, characterized by a high probability for smaller values and a rapidly decreasing probability for larger values. Finally, Fig. g illustrates a user-defined arbitrary distribution, highlighting the flexibility of the TRNG in generating custom probability profiles. This configurability is critical for integrating the TRNG into a wide range of applications, from stochastic optimization to artificial intelligence, where diverse probability distributions are needed to guide decision-making processes.

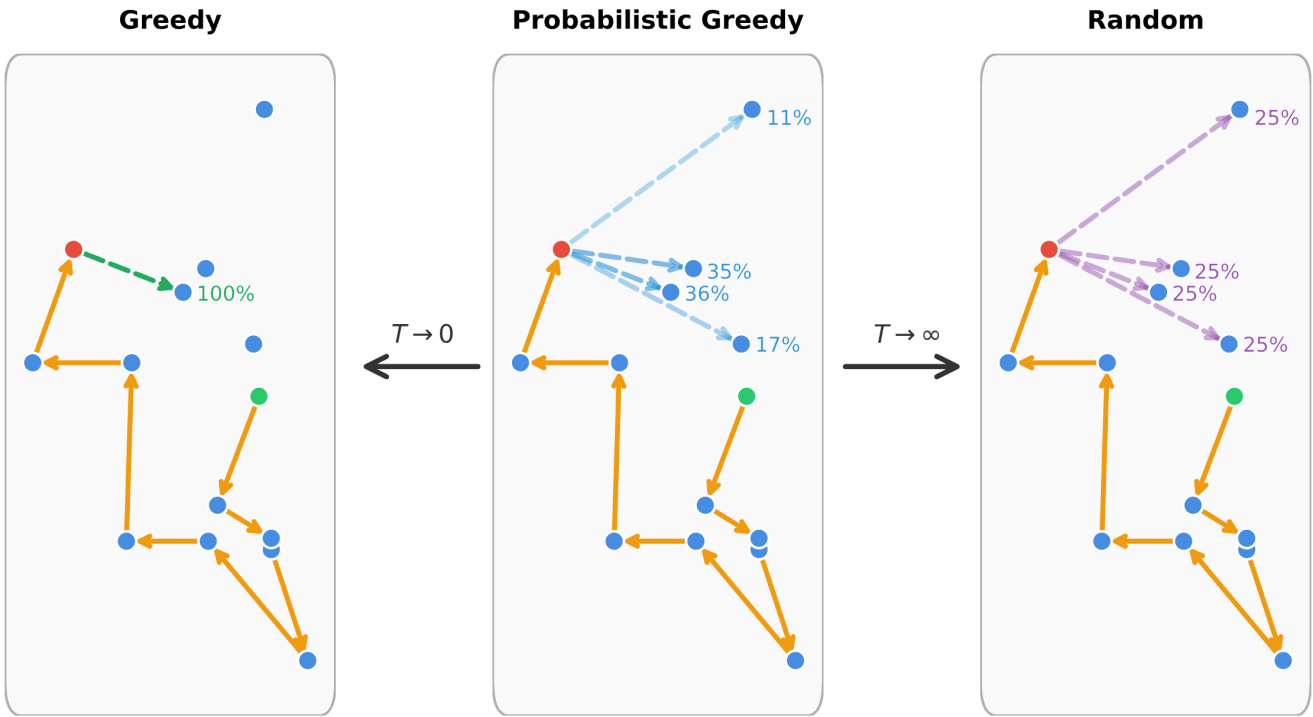

## Figure 6

When approaches 0, the algorithm always selects the closest city, equivalent to a greedy algorithm. When approaches infinity, the selection probability for each remaining city is equal, which is analogous to random selection. When is within a suitable intermediate range, a probabilistic greedy algorithm can be achieved.

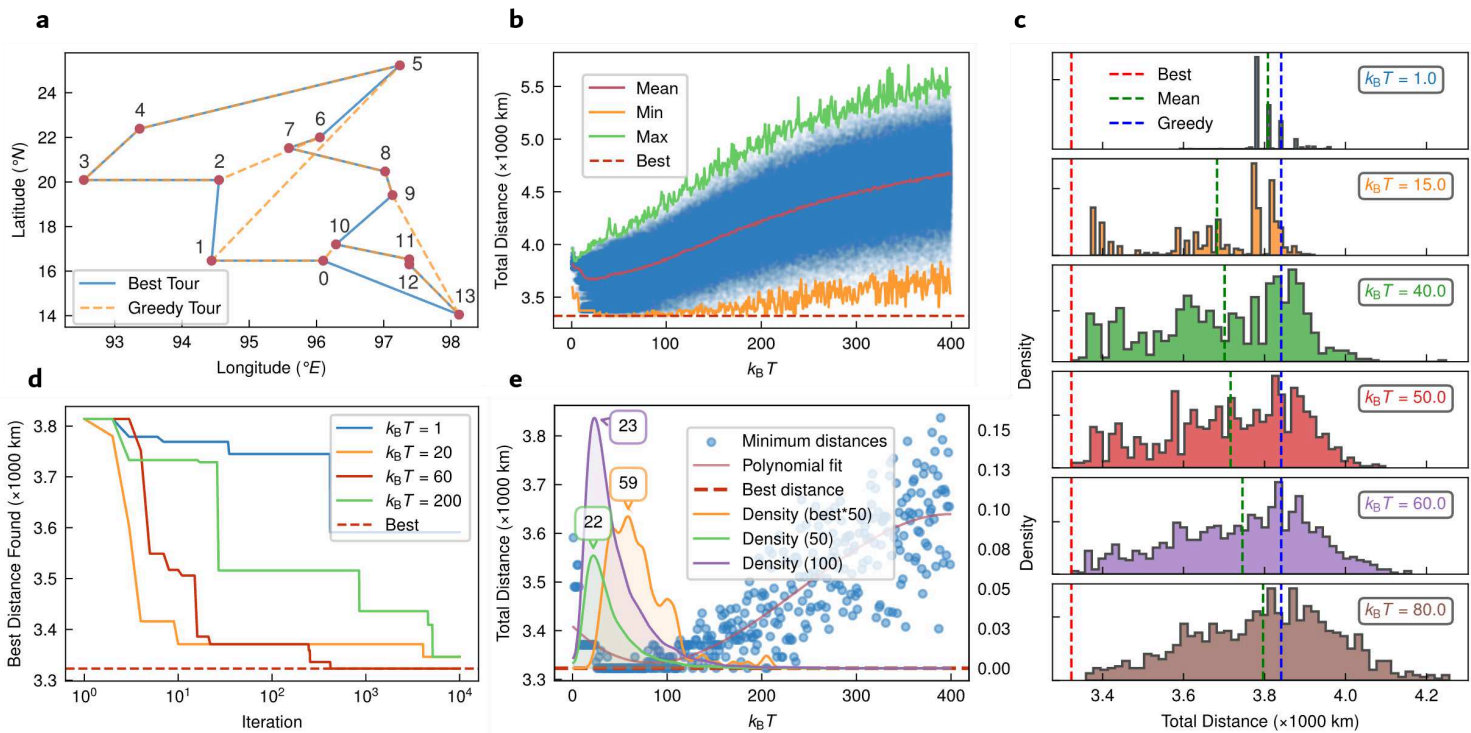

**Figure 7**

, Map of the Burma14 problem, where the solid line represents the known optimal solution and the dashed line indicates the best solution obtained using the classic greedy algorithm. , Total distance statistics of TSP solutions across the range of = 1 to 400. The orange dashed line marks the known optimal solution, while the red, green, and orange solid lines connect the maximum, minimum, and average total distances, respectively, obtained at each value. , Distribution of solution distances at six selected values, showing improved performance and reaching the optimal solution when is in the range of 40 to 60. , Relationship between the best path distance and the number of repetitions for four selected values. When = 60, the optimal solution can be achieved within 1000 repetitions. , Scatter plot of the best solutions across = 1 to 400 (left) and density distribution plots of solutions within 0, 50, and 100 kilometers of the known best solution (right).

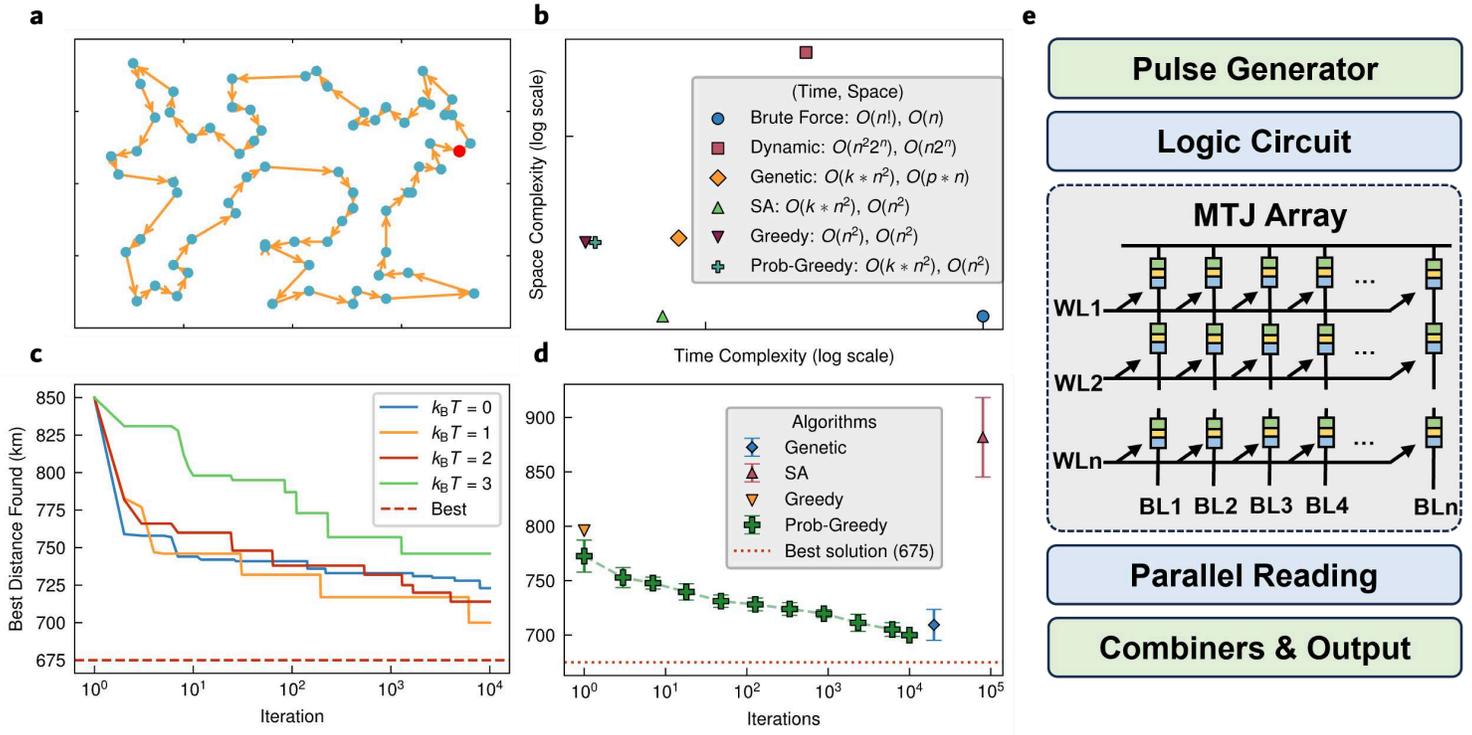

#### Figure 8

, Map of the st70 problem, where the solid line represents the known optimal solution. , A comparison of time complexity and space complexity among Brute Force, Dynamic Programming algorithm, Genetic algorithm, Simulated Annealing (SA) algorithm, Greedy algorithm, and Probabilistic Greedy algorithm, using values for  = 70. , Illustration of how the best path found in the st70 problem decreases as the number of iterations increases for = 0.1, 1.3, 2.0, and 3.0. , Comparison of solutions found by heuristic algorithms (Genetic algorithm, SA algorithm, Greedy algorithm, and Probabilistic Greedy algorithm), with the horizontal axis representing the number of iterations. , A schematic diagram of the core design of a TSP solver based on an MTJ array, highlighting the key components and connections within the solver architecture.

## Supplementary Files

This is a list of supplementary files associated with this preprint. Click to download.

- 20241210ProbGreedyTSPSupplementInformation.docx